\documentclass[preprint2]{aastex6}
\usepackage{natbib,graphicx}
\usepackage{epstopdf}                                      
\graphicspath{{./}}
\slugcomment{Accepted to ApJ}
\accepted{3 July 2017}

\newcommand{\kms}{km~s$^{-1}$}

\begin{document}

\title{{\it K2} Ultracool Dwarfs Survey II: The White Light Flare Rate of Young Brown Dwarfs}

\author{John E.\ Gizis}
\author{Rishi R. Paudel}
\author{Dermott Mullan}
\affil{Department of Physics and Astronomy, University of Delaware, Newark, DE 19716, USA}
\author{Sarah J.\ Schmidt}
\affil{Leibniz-Institute for Astrophysics Potsdam (AIP), An der Sternwarte 16, 14482, Potsdam, Germany}
\author{Adam J.\ Burgasser}
\affil{Center for Astrophysics and Space Science, University of California San Diego, La Jolla, CA 92093, USA}
\author{Peter K.\ G.\ Williams}
\affil{Harvard-Smithsonian Center for Astrophysics, 60 Garden Street, Cambridge, MA 02138, USA}

\begin{abstract}
We use {\it Kepler} {\it K2} Campaign 4 short-cadence (one-minute) photometry to measure white light flares in the young, moving group brown dwarfs 2MASS J03350208+2342356  (2M0335+23) and 
2MASS J03552337+1133437 (2M0355+11), and report on long-cadence (thirty-minute) photometry of 
a superflare in the Pleiades M8 brown dwarf CFHT-PL-17. The rotation period (5.24 hr) and projected rotational velocity ($45$ km s$^{-1}$) confirm 2M0335+23 is inflated ($R \ge 0.20 R_\odot$) as predicted for a $0.06M_\odot$, 26-Myr old brown dwarf $\beta $Pic moving group member. We detect 22 white light flares on 2M0335+23. The flare frequency distribution follows a power-law distribution with slope $-\alpha = -1.8 \pm 0.2$ over the range $10^{31}$ to $10^{33}$ erg. This slope is similar to that observed in the Sun and warmer flare stars, and is consistent with lower energy flares in previous work on M6-M8 very-low-mass stars; taken the two datasets together, the flare frequency distribution for ultracool dwarfs is a power law over 4.3 orders of magnitude.  The superflare ($2.6\times10^{34}$ erg) on CFHT-PL-17 shows higher energy flares are possible. We detect no flares down to a limit of  $2 \times 10^{30}$ erg in the nearby 
L$5\gamma$ AB Dor Moving Group brown dwarf 2M0355+11, consistent with the view that fast magnetic reconnection is suppressed in cool atmospheres. We discuss two multi-peaked flares observed in 2M0335+23, and argue that these complex flares can be understood as sympathetic flares, in which a fast-mode MHD waves similar to EUV waves in the Sun trigger magnetic reconnection in different active regions.
\end{abstract}


\section{Introduction\label{intro}}

The paradigm for the evolution of magnetic activity in low-mass main sequence stars is that magnetic braking causes the initially rapid rotation from pre-main sequence contraction to gradually decline, and this in turn causes the magnetic fields generated by the dynamo to weaken \citep{2009ARA&A..47..333D}. As a result, both the rotation rate and magnetic activity such as flaring decrease with age \citep{Gershberg:2005lr,2005ApJ...622..653T}. For fully convective $0.3 M_\odot$ stars, half the angular momentum is shed between 3 Myr {\bf and the Pleiades age} \citep{2016AJ....152..115S}. The rotation and magnetic activity evolution of brown dwarfs is quite different. Measurements of $v \sin i$ for field brown dwarfs \citep{2006ApJ...647.1405Z,2008ApJ...684.1390R} imply a mean rotation period of 4.1 hours \citep{2014ApJ...793...75R}, and a large sample of mid-infrared photometric periods confirm this view \cite{2015ApJ...799..154M}. All of these are rapid rotators compared to field stars. \citet{2009Natur.457..167C} show that turbulent dynamos can generate magnetic fields in stars, brown dwarfs and planets, and that provided the object is rapidly rotating, the strength of the magnetic fields is determined by the energy flux. \citet{2010A&A...522A..13R} show this theory implies that massive brown dwarfs have fields of a few kilogauss in their first few hundred million years, weakening to fields of 100-1000G by an age of $10^{10}$ years.  Simulations of the turbulent dynamo in fully convective stars show that both large-scale dipole and small-scale magnetic fields are generated \citep{2015ApJ...813L..31Y}. Overall, the expectation is that all brown dwarfs have significant magnetic fields, and indeed radio observations support the existence of magnetic fields even in cool T-type brown dwarfs  \citep{2012ApJ...747L..22R,2015ApJ...808..189W,2016ApJ...830...85R}.  

Despite the presence of strong magnetic fields, the fraction of a star or brown dwarf's energy converted into chromospheric activity weakens for ``ultracool dwarfs" with temperatures below 
the M6 spectral type \citep{2000AJ....120.1085G,2015AJ....149..158S}. \citet{Mohanty:2002lr} have shown this can be understood as a consequence of the increasingly neutral atmospheres: As the ionization fraction drops and the resistance increases, the magnetic fields become decoupled from the matter. These were equilibrium calculations, and as \citet{Mohanty:2002lr} noted, the existence of flares implies transient, time-dependent processes are important.  A transition from fast magnetic reconnection at high temperatures to a high resistivity regime where only slow magnetic reconnection is allowed may explain the decline in chromospheric and coronal activity but continued radio emission \citep{2010ApJ...721.1034M}. This scenario sees the fast reconnection events resulting in a range of energy release events, from many nanoflares that heat the chromosphere and corona to rarer but more powerful white light flares that can be individually observed. 
Additional parameters seem to be important in magnetic activity: Magnetic topology may explain the difference between radio-quiet, X-ray bright  dwarfs and the radio-loud, X-ray faint dwarfs \citep{2014ApJ...785...10C,2014ApJ...785....9W}.  

Even setting aside the numerous radio-only bursts, it is well established that X-ray and optical flares do occur in stellar late-M and early-L dwarfs -- some notable examples include the very first optical spectrum of \object{VB10}, the first known M8 dwarf \citep{1956PASP...68..531H}, the discovery of a nearby M9 dwarf due to a huge X-Ray flare with $L_X/L_{\rm bol} =0.1$ \citep{Hambaryan:2004mz}, and an L0 dwarf with a $\Delta V<-11$ white light flare \citep{2016ApJ...828L..22S}. {\bf The serendipitous optical spectra of the M7-M9 dwarf flares reported by \citet{Bessell:1991rw},  \citet{1999MNRAS.302...59M}, \citet{1999ApJ...519..345L}, \citet{1999ApJ...527L.105R}, \citet{2000AJ....120.1085G}, and \citet{Martin:2001qy} all showed strong atomic emission lines and many included veiling or a blue continuum.}
A difficulty, however, with flare studies is that detectable flares are rare enough that it is difficult to assess their frequency {\bf as a function of energy}. \citet{2011PhDT.......144H} monitored four field M6-M8 stars in U-band and found 39 flares over 59 hours. These flares followed a power-law frequency distribution, as seen in hotter flare stars, but with a rate comparable to ``inactive" but more luminous M0-M2.5 dwarfs.  Similarly, {\it Kepler} optical monitoring of a  L1 dwarf star also found a power-law flare frequency distribution \citep{2013ApJ...779..172G}. Young M-type brown dwarfs with similar $T_{\rm eff}$ also exhibit flares, such as the $\sim500$-Myr M9 lithium brown dwarf \object{LP 944-20} \citep{2000ApJ...538L.141R}. X-Ray flares, as well as quiescent emission, have also been reported from very young ($<5$ Myr) M6-M9 brown dwarfs in Orion \citep{2005ApJS..160..582P} and Taurus \citep{Grosso:2007vn}.

By monitoring over 100,000 stars over four years, {\it Kepler} mission \citep{2010ApJ...713L..79K} is known to have detected over 800,000 flares in 4041 stars \citep{2016ApJ...829...23D}.  These include superflares in A and F stars with thin convective zones  \citep{2012MNRAS.423.3420B,2015MNRAS.447.2714B} and solar-like G dwarfs \citep{2012Natur.485..478M} as well as fully-convective M dwarfs {\bf \citep{2013A&A...555A.108M,2014ApJ...797..121H}} and even an L1 dwarf \citep{2013ApJ...779..172G}. These flares are detected by white light emission enhancing the normal stellar photospheric emission through {\it Kepler}'s broad (430nm - 900nm) filter; thus, only extremely energetic events ($>10^{34}$ erg) are seen in the warmer stars but weaker flares can be seen in the coolest stars.  
In the models of \citet{2015SoPh..290.3487K}, a beam of non-thermal electrons with a energy flux of $10^{13}$ erg cm$^{-2}$ s$^{-1}$ can produce a dense, hot chromospheric condensation that emits white light like a $\sim10,000$K blackbody. Solar flares with energies $\sim10^{31}$ erg also emit most of their energy in white light like $\sim9000$K blackbodies \citep{2011A&A...530A..84K} --- flares of this energy would not be detectable by {\it Kepler} on solar-type stars but were detected on the L1 dwarf and the M4 dwarf GJ 1243 \citep{2016ApJ...829..129S}.  

The extended {\it Kepler} {\it K2} mission \citep{2014PASP..126..398H} allows many new targets to be observed. We are using {\it K2} to monitor ultracool dwarfs for white light flares as well as measuring rotational periods and searching for transits. Targets that happen to lie within each {\it K2} field of view are monitored for a $\sim2.5$ month-long campaign.  The overall aim of our survey is to measure quantities such as the flare frequency, maximum flare energy, flare light curve morphology in order to understand their dependence on parameters such as temperature, mass, and age. In this paper, we present {\it K2} Campaign 4 observations of three brown dwarfs which are confirmed members of  nearby moving groups and clusters, so that unlike
most field dwarfs, their age, mass, radius and other parameters are well determined.  We present the target properties in Section~\ref{s:targets}, the {\it K2} observations in in Section~\ref{s:K2data},  and discussion of the magnetic activity in Section~\ref{s:discussion}.



\section{Targets and Spectroscopy}\label{s:targets}

\subsection{Target Characteristics}

In Table~\ref{tab:prop}, we list the key properties of our targets.\footnote{EPIC 211046195 and EPIC 210327027 were observed for K2 Guest Observer Program  4036 (PI Gizis); EPIC 211110493 was observed for GO Programs 4024 (PI Lodieu), 4026 (PI Scholz), and 4081 (PI Demory).}  
\object{2MASS J03350208+2342356} (hereafter 2M0335+23) was observed by {\it K2} as source EPIC 211046195.
This field ultracool dwarf was discovered by \citet{2000AJ....120.1085G}, who classified it as M8.5 in the optical and noted H$\alpha$ emission. {\bf It is apparently single in {\it Hubble Space Telescope} imaging \citep{2003AJ....125.3302G}.}
  \citet{hiresmdwarfs} detected significant rotational broadening ($v \sin i \approx$ 30 \kms), lithium in absorption, and again detected H$\alpha$ emission. The presence of lithium identified this object unambiguously as a brown dwarf {\bf \citep{1992ApJ...389L..83R}.}  \citet{2012ApJ...758...56S} measured its trigonometric parallax distance to be $42.4 \pm 2.3$ pc and showed that its distance and space velocity identified it as a member of the $\beta$ Pic Moving Group (BPMG).  The latest parallax from \citet{2016ApJ...833...96L} places it at $46\pm 4$ pc; for consistency with the literature we adopt the nominally more precise distance of 42.4pc for our analysis.   The age of this group is 
$24 \pm 3$ Myr \citep{2014MNRAS.445.2169M,2015MNRAS.454..593B}.  \citet{2015ApJS..219...33G} have analyzed 2M0335+23 in detail to find that it is $60.9^{+4.0}_{-4.4}$ jupiter masses (i.e, $0.058\pm0.004 M_\odot$) with a radius of $2.40 \pm 0.04$ jupiter radius according to models \citep{2015ApJS..219...33G}.  Adopting BC$_J = 2.0$ \citep{2015ApJ...810..158F}, the luminosity is $10^{-2.55} L_\odot$ and $T_{\rm eff} = 2700$K. \citet{2014A&A...563A..45S} measured it to have apparent (AB) magnitude $i=15.601$ (as source DANCe 5121623). Infrared spectroscopy confirms that it is lower surface gravity than ordinary field dwarfs, with a classification of M7 VL-G  \citep{2013ApJ...772...79A} and M7.5$\beta$ \citep{2015ApJS..219...33G}. In the optical, low surface gravity leads to enhanced VO features \citep{2008ApJ...689.1295K}; this would bias \citet{2000AJ....120.1085G}'s classification to a later type. We re-classify the spectrum as M7$\beta$ in the optical. In particular, the optical spectrum of 2M0335+23 is definitely ``earlier" (warmer) than young M9 brown dwarf 2MASS J06085283-2753583 \citep{2003AJ....126.2421C} {\bf which may be a BPMG member \citep{2010ApJ...715L.165R} or more likely a 40-Myr-old Columba member \citep{2016ApJ...833...96L}.}  The observed H$\alpha$ emission line strength of EW$\approx 5\AA$ imply $\log H\alpha/L_{bol} \approx -5.5$ \citep{2014PASP..126..642S}. Despite 2M0335+23's rapid rotation and youth, this places it in the bottom half of the M7 activity range \citep{2014arXiv1410.0014S}. Finally, we use
the mass, luminosity and radius to predict theoretical mean surface magnetic fields of our targets using Equation 1 of \citet{2010A&A...522A..13R}.  

\object{2MASS J03552337+1133437} (hereafter 2M0355+11) was observed by {\it K2} as source EPIC 210327027.  
Discovered by \citet{Reid:2006fj} and classified as low-surface gravity (L5$\gamma$) {\bf with lithium} by \citet{2009AJ....137.3345C}, this brown dwarf is now recognized as a dusty, low-surface-gravity member of the AB Doradus Moving Group (ABDMG) which shares many spectral characteristics with directly imaged exoplanets \citep{Faherty:2013qy,Liu:2013lr}. \citet{2015MNRAS.454..593B} derive an ABDMG age of $149^{+51}_{-19}$ {Myr}; note that this age is tied to the Pleiades age of $130 \pm 20$ Myr from lithium depletion \citep{2004ApJ...614..386B} {\bf that has been updated to $112 \pm 5$ Myr \citep{2015ApJ...813..108D}}. \citet{2016ApJ...833...96L} measure a trigonometric parallax of {\bf $109.5\pm 1.4$ mas}, and we use a distance of 9.1 pc for the rest of this paper. For discussion purposes, we adopt the values derived by  \citet{2015ApJ...810..158F}:  $\log L/L_{bol} = -4.10 \pm 0.03$, radius $R=1.32\pm0.09 R_J$, surface gravity $\log g = 4.45 \pm 0.21$, $T_{\rm eff} = 1478 \pm 57$K, and mass $M=19.98 \pm 7.76 M_J$ (i.e., $\sim0.02 M_\odot$).  \citet{2010ApJ...723..684B} measured $v \sin i = 12.31 \pm  0.15$ \kms, noting that it is an unusually slow rotator for a L dwarf. This, however, still implies a minimum rotation period of 13 hr, a rapid rotator compared to M dwarf stars. No H$\alpha$ emission has been detected (EW$<24.29$\AA, \citealt{2007AJ....133.2258S}) though the upper limit is above the emission level of most L dwarfs.  

\object[2MASS J03430016+2443525]{CFHT-PL-17}, a brown dwarf member of the Pleiades discovered by \citet{1998A&A...336..490B}), was observed by {\it K2} as source EPIC 211110493. \citet{2000ApJ...543..299M} classified it as optical spectral type M7.9 (which we will hereafter round off to M8.) and found H$\alpha$ emission (EW$ \approx 7$\AA). \citet{2015A&A...577A.148B} confirm it has a 100\% chance of being a cluster member and measure $i=19.745$ (AB). We adopt the VLBI Pleiades distance of $136.2 \pm 1.2$ pc \citep{2014Sci...345.1029M}; \citet{2015A&A...577A.148B}, using the same distance, find a luminosity of $0.0008456 L_\odot$ ($T_{\rm eff} = 2500$K) which implies the mass is $0.06 M_\odot$.  Thus, CFHT-PL-17 is very similar in mass to 2M0335+23, but older and one spectral type later ($\sim200$K cooler), and it is a similar age to 2M0355+11, but more massive and warmer.\footnote{We also identify a flare in the candidate Pleiades brown dwarf  \object[2MASS J03454521+2258449]{BPL 76} \citep{2000MNRAS.313..347P}, observed as EPIC 211000317. \citet{2015A&A...577A.148B}, however, have measured its proper motion and assign it a membership probability of zero.}

\begin{deluxetable*}{llllrrrcrr} 
\tablewidth{0pc} \tabletypesize{\scriptsize}
\tablecaption{Key Target Properties \label{tab:prop} }
\tablehead{ 
\colhead{Object}  & \colhead{EPIC} & \colhead{$\widetilde{K}_p $} & \colhead{Type} & \colhead{Distance} & 
\colhead{Age} & \colhead{Mass} & \colhead{$\log L/L_\odot$} &\colhead{$T_{\rm eff}$} & \colhead{$B$} }
\startdata
2M0335+23 & 211046195 & 16.7 & M7$\beta$ & 42.4 pc & 24 Myr & $0.06 M_\odot$ & -2.55  &2700K &  2.2 kG \\
2M0355+11 & 210327027 & 20.4 & L5$\gamma$ & 9.1 pc & 150 Myr & $0.02 M_\odot$ &-4.10  &1480K & 1.1 kG \\
CFHT-PL-17 & 211110493 & 20.8 & M8$\beta$ & 136 pc & {\bf 112} Myr & $0.06 M_\odot$ & -3.07 &2500K & 2.5 kG \\
\enddata
\tablecomments{Parameters given are rounded off. The mean surface magnetic field $B$ is theoretical. See text for references and uncertainties. }
\end{deluxetable*}

\subsection{New Spectroscopy}

We observed 2M0335+23 on UT Date 2016 February 3 with the Keck NIRSPEC spectrograph \citep{McLean:2000lr} to obtain spectra with $\lambda / \Delta \lambda = $20,000 in the 2.3 $\mu$m region dominated by CO bands. Conditions were clear with 1\arcsec~seeing. We obtained two exposures of 750 sec each, following observation of the A0V star HD 19600 for telluric calibration. 

The setup and data analysis were as described in \citet{2013ApJ...779..172G}. We achieved a typical signal-to-noise of $>50$ for these observations. We find $v_{rad} = 12.6 \pm 1.0$ \kms and $v \sin i = 45.4 \pm 3.4$ \kms. This radial velocity increases 2M0335+23's probability of BPMG membership to 96.5\% using the \citet{2012ApJ...758...56S} astrometry in the BANYAN II model \citep{2014ApJ...783..121G,2015ApJ...798...73G}.  

\begin{figure*}
\plotone{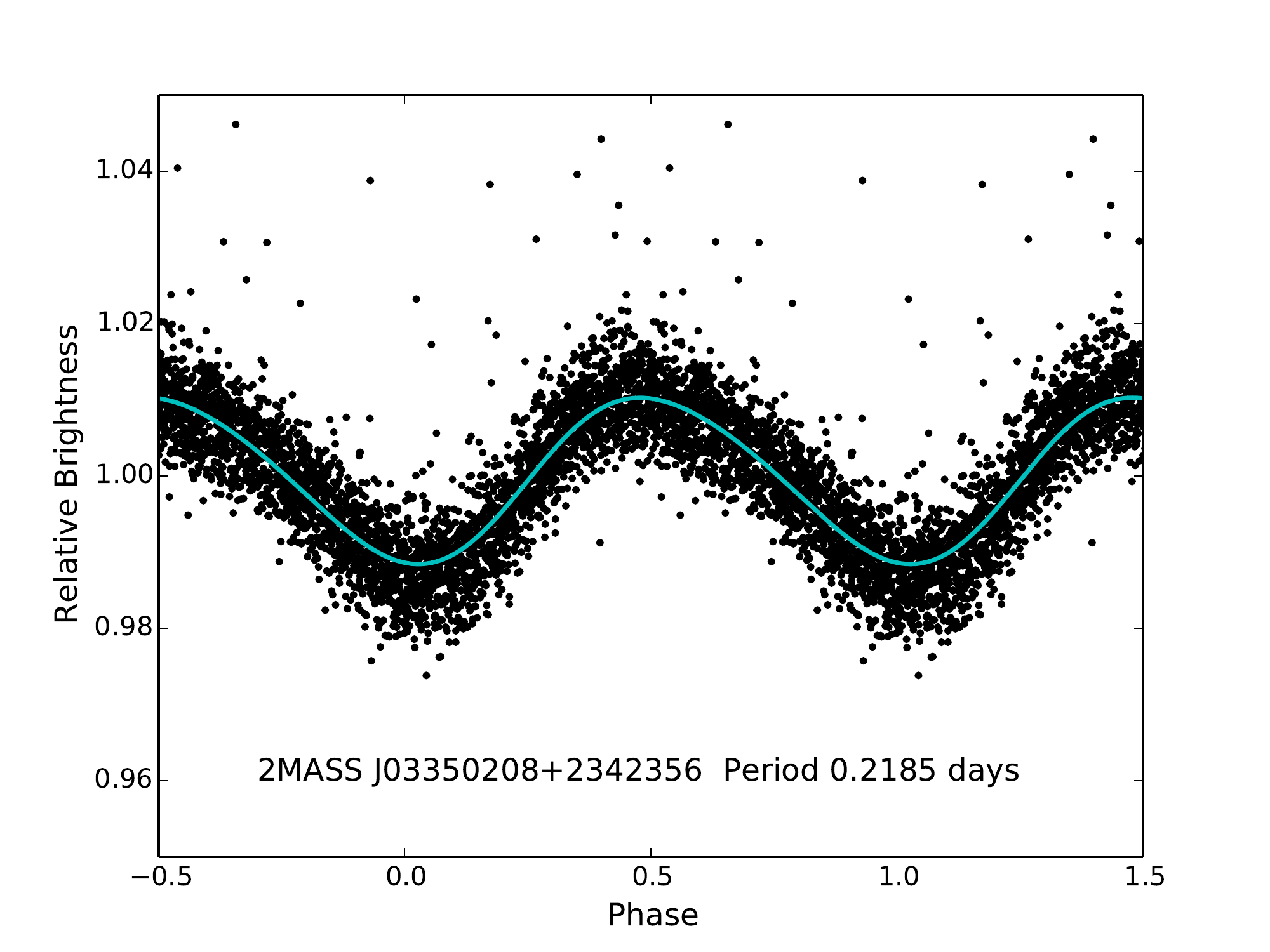}
\caption{{\it K2} long-cadence photometry for 2M0335+23 phased to a period of 0.2185 days.  The pipeline-estimated uncertainty on each point is 0.0017. All points are plotted twice, and a non-parametric regression fit with a quadratic local polynomial kernel is shown in cyan.  \label{fig-phased0335}}
\end{figure*}

\section{{\it K2} Photometry}\label{s:K2data}

{\it Kepler} records the pixels for every target as averages over ``long" (30 minute, \citealt{Jenkins:2010fk}) cadences; for 2M0335+23 and 2M0355+11, it also recorded ``short" (1 minute, \citealt{2010ApJ...713L.160G}) cadence data. We report {\it Kepler} mission times, which are equal to BJD - 2454833.0.  

The brightnesses of Kepler and {\it K2} targets are described on the $K_p$ magnitude system \citep{2011AJ....142..112B} tied to ground-based photometry; this system was not designed for ultracool dwarfs and the EPIC catalog \citep{2016ApJS..224....2H} magnitudes for our targets are not useful. \citet{2015ApJ...806...30L}  defined $\widetilde{K}_p \equiv 25.3 - 2.5 \log({\rm flux})$, where $\widetilde{K}_p \approx K_p$ for most (e.g., AFGK-type) stars and ``flux" is the count rate measured through a 3-pixel radius aperture.  We find that the apparent $\widetilde{K}_p$ magnitudes of 2M0335+23, 2M0355+11, and CHFT-PL-17 are 16.7, 20.4, and 20.8. By using $\widetilde{K}_p$ we can discuss both extremely red sources and time-dependent (blue) flares in a terms of the well established $K_p$ system.

2M0335+23 is bright enough that photospheric variability is detectable. We use the {\it K2} mission pipeline photometry corrected for the effects of pointing drift and other systematic errors.
For 2M0335+23, the Lombe-Scargle periodogram shows a strong signal at $P=0.2185$ day (5.244 hour), which we identify as the rotation period of the brown dwarf. The phased data are shown in Figure~\ref{fig-phased0335} normalized to the median. We have verified that other {\it K2} pipelines \citep{2014PASP..126..948V,2016MNRAS.456.2260A,2016MNRAS.459.2408A} give consistent results for this source. We do not detect periodic photometric variations in 2M0355+11 or CFHT-PL-17. We note that in the case of the {\it K2} mission corrected photometry for 2M0355+11, there is a periodic signal of 1.11 days, but we believe this is a spurious signal, and it is not present in the other reduction pipelines.  

We measured short cadence photometry using the {\it K2} Release 10 pixel files. 
(There is not yet any {\it K2} mission official light curve product for short cadence data.)
We used circular aperture photometry ({\rm photutils}) with radius of 2 pixels centered on the target; but we verified that our results are qualitatively unchanged with circular apertures of radius 3 or rectangular fixed apertures. The sources, especially 2M0355+11, are faint enough that centroiding introduces photometric noise; we adopt a best position based on the median of all centroid measurements, and then adjust it  for each observation using the spacecraft motion estimate calculated by the mission (recorded as POS\_CORR1 and POSS\_CORR2 in the FITS file headers). The photometry shows the usual systematic drifts, but these have little effect on measurements of flares which have timescales of a few minutes. 

\begin{figure*}
\plotone{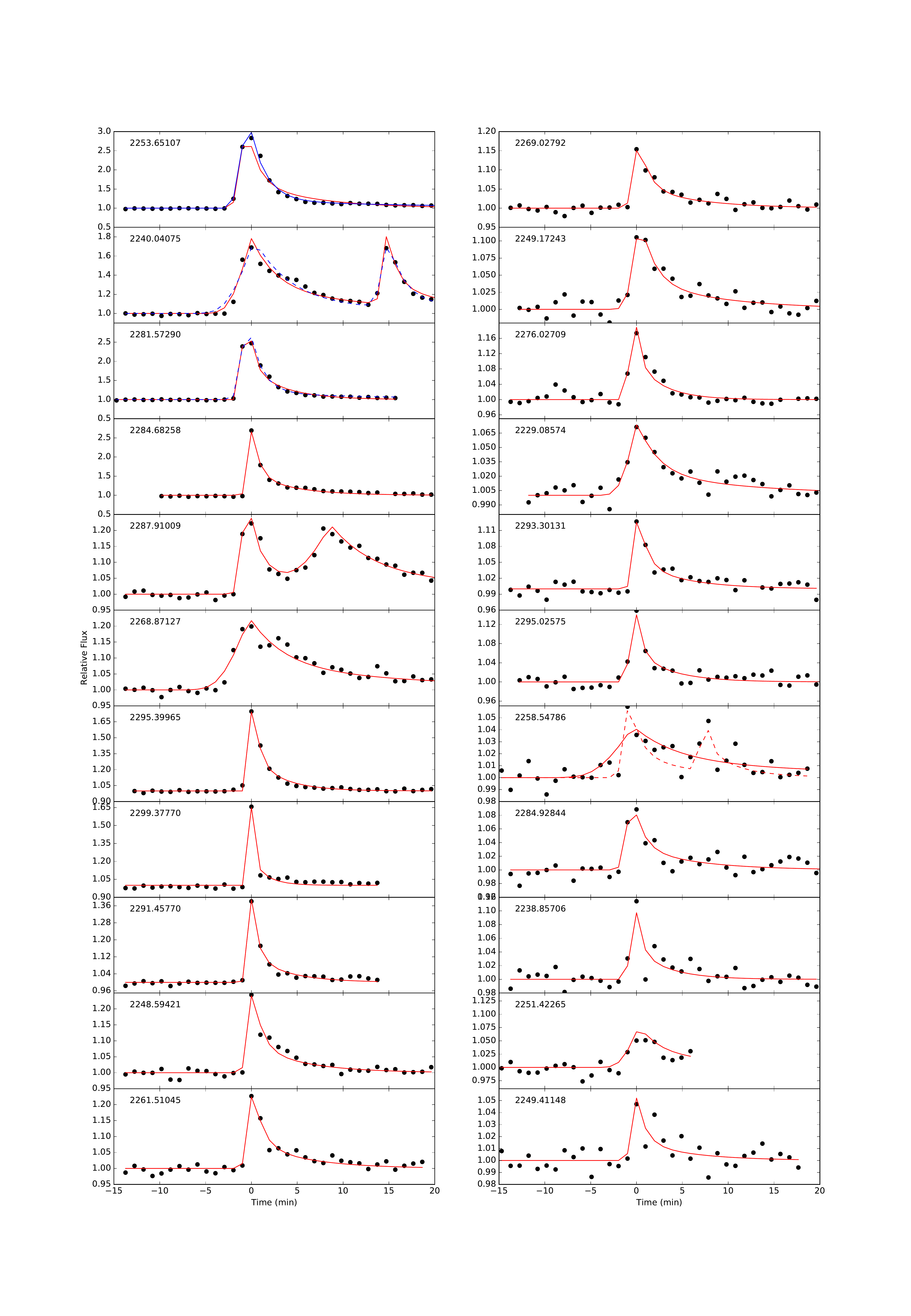}
\caption{Photometry of each detected flare on the brown dwarf 2M0335+23. Model light curve fits using the \citet{2014ApJ...797..122D} (D14) {\it Kepler} M dwarf flare template are shown in red. An alternative model fit to the  flare at time 2253.65107 is shown in blue; the same parameters are shown in dashed blue for the next two strongest flares as well. For the faint flare at mission time 2258.54786 we plot the results of fitting both a single (solid red) and double (dashed red) flare D14 templates.   \label{fig-flares}}
\end{figure*}

We lack short cadence photometry for CFHT-PL-17, but motivated by strong flare at mission day 2261.94 which we noticed in the mission pipeline photometry, we measure our own 2 pixel radius long cadence photometry in the same way to analyze its flare.  
We identified 22 flares in 2M0335+23 by visually examining the lightcurves; they are shown in Figure~\ref{fig-flares} and listed in Table~\ref{tab:flares}.  We reject all events that brighten in only a single observation or are not centered on the target's position. No flares are detected in 2M0355+11; we note however that a passing asteroid creates a spurious brightening in aperture photometry at mission time 2271.13.  

For each 2M0335+23 flare, we fit the M dwarf flare light curve described by \citet{2014ApJ...797..122D}, hereafter D14, who found that most flares on the M4 dwarf GJ 1243 could be described by a fast rise (a 4th order polynomial) and a slower double exponential decay:

\begin{equation}
\Delta F =  A (\alpha_i e^{-\gamma_i \Delta t/t_{1/2}} + \alpha_g e^{-\gamma_g \Delta t/t_{1/2}})
\label{D14equation}
\end{equation}

All flare light curves are then described by nine universal parameters: 
$\alpha_i=0.6890 (\pm 0.0008)$, $\alpha_g = 0.3030 (\pm 0.0009)$, $\gamma_i=1.600(\pm0.003)$, $\gamma_g = 0.2783(\pm0.0007)$ plus the five polynomial parameters given in D14. Each flare also has three unique parameters: the peak amplitude of the flare, the full-width time at half-max  ($t_{1/2}$), and the time of the flare peak. We fit four free parameters to each flare using {\rm emcee} \citep{2013PASP..125..306F}: the non-flaring photosphere, the peak amplitude of the flare, the full-width time at half-max  ($t_{1/2}$), and the time of the flare peak. The fits are shown in red in Figure~\ref{fig-flares}.  
Two of the brighter flares (on mission days 2240 and 2287) are complex flares with two peaks in their light curves: We have fit them as the superposition of two flares. 
The equivalent duration listed in Table~\ref{tab:flares} is a measure of the flare energy
compared to the quiescent luminosity; it is obtained by integrating the observed flare count rate and dividing by the photosphere count rate.  This is a distance-independent measure of the flare, but it is dependent on the filter and the photospheric properties of the stars:  Our durations will be much longer than otherwise identical flares observed on a G dwarf due to the lower photospheric flux.  Our detection of weak flares is limited by noise, and it is clear that we cannot reliably detect flares below equivalent durations of 20s. The flares on mission days 2238, 2249, and 2251 in Figure~\ref{fig-flares} are examples of marginal detections that may be some form of correlated noise rather than real flares. The flare on mission day 2299 is also questionable because most of the flux occurs in a single time period; its exclusion would have negligible effects on the remaining analysis.

The strongest 2M0335+23 flare (mission time 2253.65107) has noticeable deviations from the D14 template: The best fitting model (red) under-predicts the peak and over-predicts the gradual phase, with a $t_{1/2}$ that is too short. We therefore use a new model template in which we keep the polynomial rise parameters fixed but allow the decay parameters to be fitted. This adds three free parameters, since we require ($\alpha_i + \alpha_g = 1$). The results are: $\alpha_i=0.9233\pm 0.0055$ ($\alpha_g=0.0767$), $\gamma_i = 1.3722 \pm 0.054$, and $\gamma_g=0.1163 \pm 0.011$. \citet{2014ARep...58...98G} argues during the impulsive decay phase cooling is by blackbody emission, which suggests the relative contribution of this component of radiative cooling was different in this flare.  

\begin{figure*}
\plotone{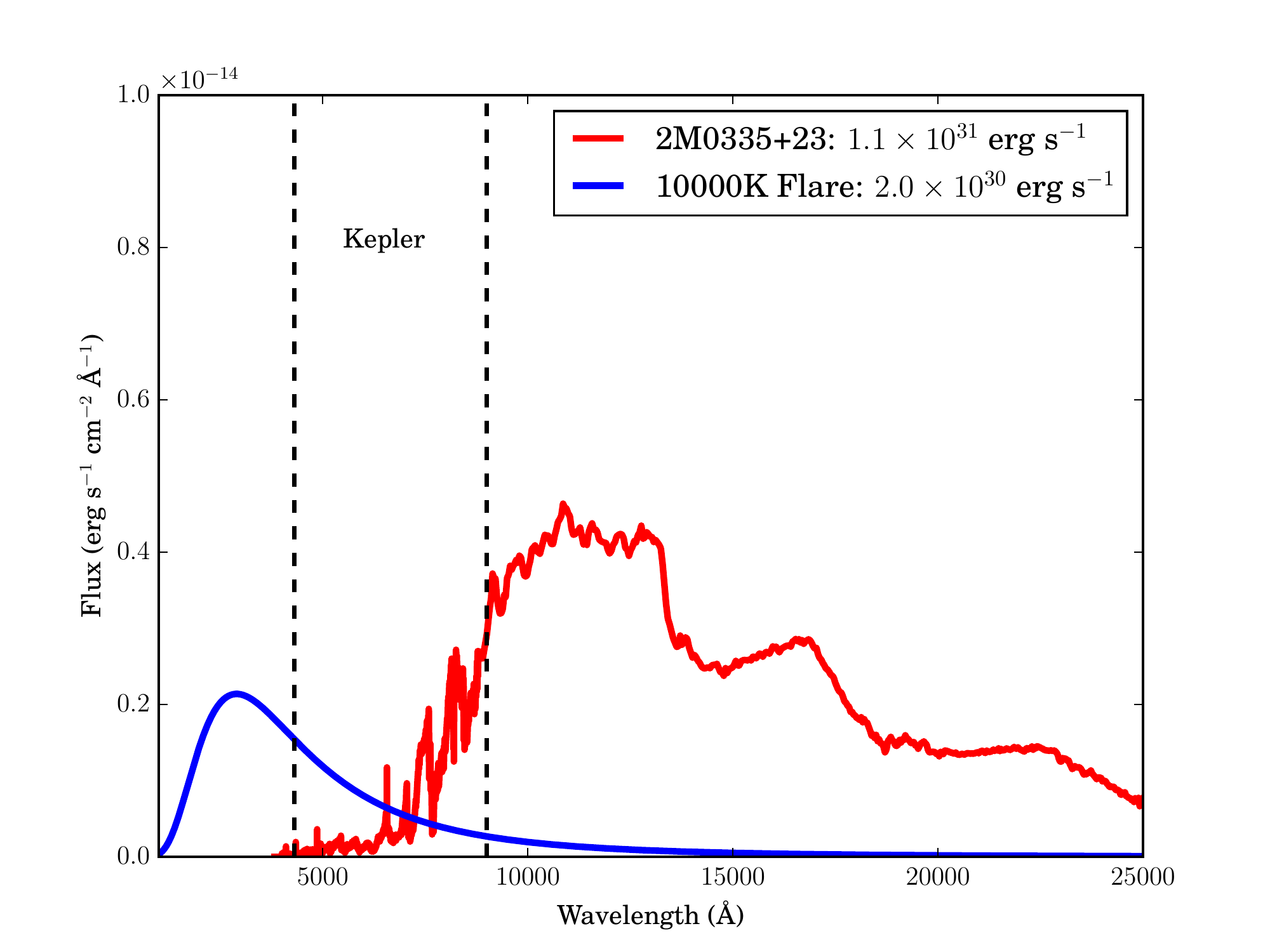}
\caption{The optical and near-infrared spectral energy distribution of 2M0335+23 in red with a hypthetical 10,000K blackbody flare that produces the same {\it Kepler} count rate.  The infrared spectrum is from \citet{2013ApJ...772...79A} and optical is an M7 standard \citep{Bochanski:2007ys} scaled to the photometry. 
Using the distance of 42.4 pc,  the bolometric luminsoity of 2M0335+23 is $1.1 \times 10^{31}$ erg s$^{-1}$, mostly emitted longwards of the broad {\it Kepler} optical filter, and the flare is $2.0 \times 10^{30}$ erg s$^{-1}$, much emitted shortwards.  \label{fig-calib}}
\end{figure*}

To calibrate equivalent duration in terms of energy, we follow the procedures described in \citet{2013ApJ...779..172G}. Because flares are much hotter than the brown dwarf targets, white light flares have a higher average energy per detected {\it K2} photon than the photosphere. The photosphere is modeled with an active M7 dwarf template \citep{Bochanski:2007ys} scaled to the measured $i$ photometry and known trigonometric parallax distance. The flare is modeled as an 10,000K blackbody, which gives good agreement with the flare measurements in \citet{Hawley:1991uq}. Figure~\ref{fig-calib} shows the optical and near-infrared spectral energy distribution of 2M0335+23 and flare with the same count rate through the {\it Kepler} filter.
We find that the a flare with equivalent duration of 1 second has a total (bolometric UV/Vis/IR) energy of $2.0 \times 10^{30}$ erg. We emphasize that we have extrapolated to wavelengths not detected by {\it K2} and
that our analysis includes atomic emission features between 430nm and 900nm in the observed ``white light" photometry. Finally, we also report the peak (short cadence) absolute $\widetilde{K}_p$ magnitude of the flare.  For these, we have applied an aperture correction of 1.08 to correct the $r=2$ pixel aperture to the $r=3$ pixel aperture.  We note that the total equivalent duration of detected flares is 0.030 days, so that just 0.04\% of the 2M0335+23's optical light over the course of the campaign is due to white light flares.

\begin{deluxetable*}{rrrrrl} 
\tablewidth{0pc} \tabletypesize{\scriptsize}
\tablecaption{Flares in 2M0335+23 \label{tab:flares} }
\tablehead{ 
\colhead{Time}  & \colhead{Peak Ratio} & \colhead{Eq. Duration} & \colhead{Energy} & \colhead{Secondary Time} & \colhead{Comments} \\
\colhead{(d)} & & \colhead{(s)} & \colhead{($10^{31}$ erg)} & \colhead{(d)}  
 }
\startdata
2253.65107 & 1.83 & 518 & 103.5 & \nodata \\
2240.04075 & 0.68 & 494 & 98.9 & 2240.0511 & Complex\\
2281.57290 & 1.47 & 339 & 67.8 & \nodata \\
2284.68258 & 1.69 & 267 & 53.4 & \nodata \\ 
2287.91009 & 0.22 & 203 & 40.5 & 2287.9161 & Complex\\
2268.87127 & 0.20 & 145 & 29.0 & \nodata  \\ 
2295.39965 & 0.75 & 110 & 22.0 & \nodata \\ 
2299.37770 & 0.66 & 56 & 11.2 & \nodata & Questionable \\ 
2291.45770 & 0.38 & 53 & 10.7 & \nodata \\ 
2248.59421 & 0.24 & 48 & 9.6 & \nodata \\ 
2261.51045 & 0.23 & 47 & 9.4 & \nodata \\ 
2269.02792 & 0.15 & 35 & 7.0 & \nodata \\ 
2249.17243 & 0.11 & 35 & 6.9 & \nodata \\ 
2276.02709 & 0.17 & 32 & 6.4 &\nodata \\  
2229.08574 & 0.07 & 28 & 5.5 & \nodata \\  
2293.30131 & 0.13 & 26 & 5.2 & \nodata \\ 
2295.02575 & 0.15 & 25 & 5.0 & \nodata \\ 
2258.54786 & 0.04 & 24 & 4.8 & \nodata & Complex? \\ 
2284.92844 & 0.09 & 23 & 4.7 & \nodata \\ 
2238.85706 & 0.11 & 18 & 3.5 & \nodata & Questionable \\  
2251.42265 & 0.05 & 13 & 2.6 &\nodata & Questionable\\ 
2249.41148 & 0.05 & 10 & 1.9 & \nodata & Questionable\\ 
\enddata
\end{deluxetable*}

We detect no flares in 2M0355+11. We verified that we could have detected flares by adding our observed 2M0335+24 flare data back into the 2M0355+11 at random times, and recovered them all. Because 2M0355+11 is 4.7 times closer than 2M0335+24, a flare with 
22 times less energy would produce the same count rate.  We conclude that we could have detected flares with $E>2.0\times10^{30}$ ergs, and place a 95\% confidence upper limit of 3 such flares over 70.7 days.  (The effect of 2M0355+11 photosphere's much fainter apparent magnitude would simply to be increase the equivalent duration or relative amplitude of the flare.) If the timescale for these flares, however, was less than one minute than we could not distinguish them from cosmic rays or other noise sources. However, because the late-M dwarf flares (\citealt{2011PhDT.......144H}, Table 4.1) with energies at or above our limit have timescales of several minutes, we conclude that this effect is not a concern.  

\begin{figure*}
\plotone{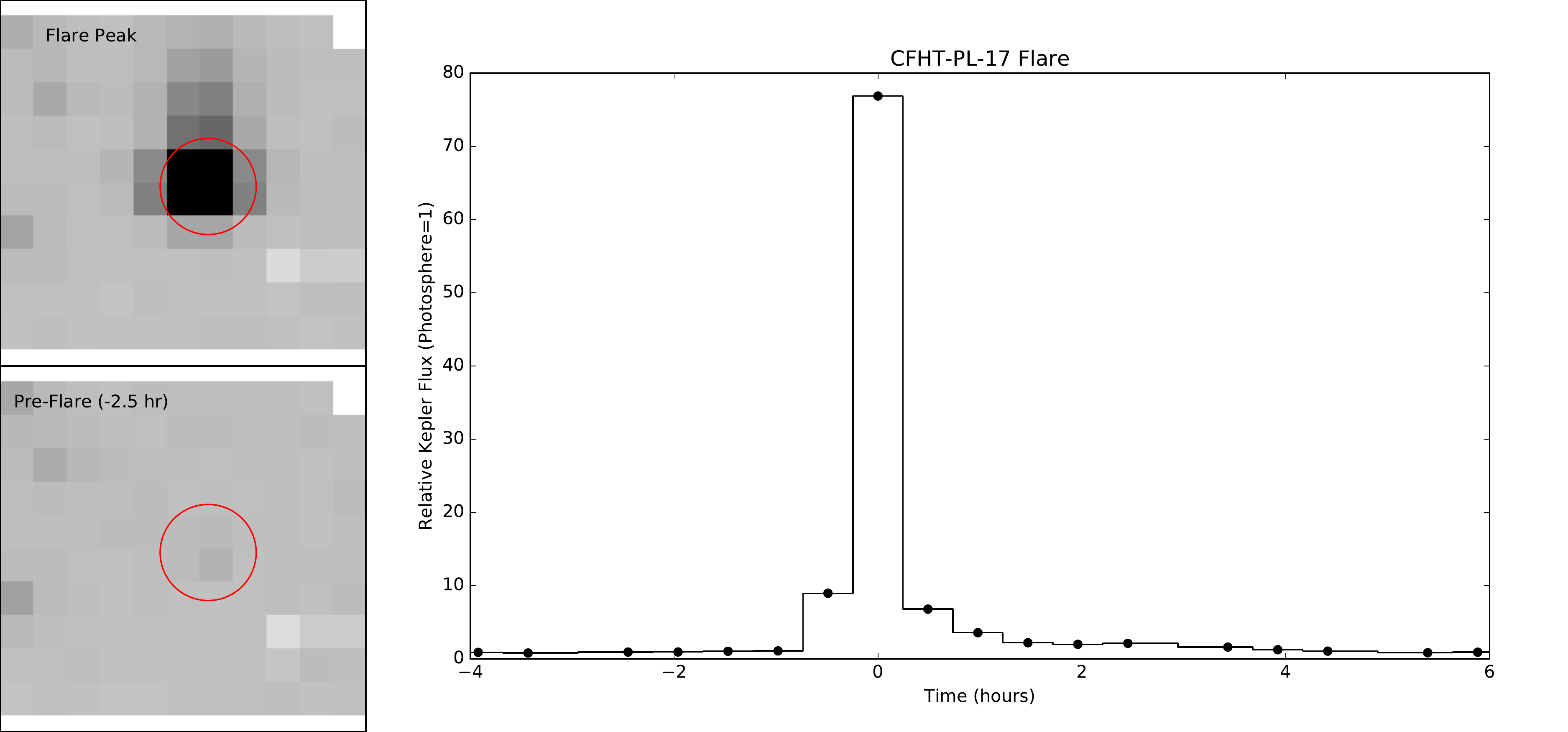}
\caption{{\it K2} images of the Pleiades brown dwarf CFHT-PL-17 at flare peak (top left) and 2.5 hours before (bottom left) and the light curve of the flare (right). Each photometric data point is the average brightness over a 30 minute window. The photometry peaks at 77 times the non-flaring photosphere at mission time 2261.9612.  If sampled on short cadence, the peak would have been considerably brighter.  \label{fig-superflare}}
\end{figure*}

We measure the flare on CFHT-PL-17 using 3-pixel radius aperture photometry (Figure~\ref{fig-superflare}).  
The flare is first detectable at mission time 2261.9407 where it has brightened to 9.0 times the original photosphere. In the 2261.9612 exposure, it has reached 77 times the photosphere to 
achieve $\widetilde{K}_p= 16.0$. 
The flare then declines, with the last detectable excess of 24\% at 2262.1246, for a total observed duration of five hours.  The equivalent duration is 170,000 s (2.0 days). Using the same calibration procedure with an active 
M8 template \citep{Bochanski:2007ys}, we calculate the flare energy is $2.6 \times 10^{34}$ erg. Given the sharply peaked light curve, we conclude that $t_{1/2} < 30$ min.   

We can fit the D14 template by computing it on one-minute timesteps but comparing to the long-cadence data, as in our analysis of an L dwarf super flare in Paper I \citep{2017ApJ...838...22G}. We find that the peak is 380 times the photosphere ($\widetilde{K}_p \approx 14.3$), with $t_{1/2}$ of 3.9 minutes. This should be viewed with caution because we do not know if the D14 template applies to this flare, or if the flare was complex and multi-peaked. On the other hand, {\it Kepler} short cadence photometry of comparable energy flares in F stars shown by \citet{2012MNRAS.423.3420B} are sharply peaked.

\section{Discussion}\label{s:discussion}

\subsection{The Radius of a 24 Myr Brown Dwarf}

Young brown dwarfs should have inflated radii compared to field stars of the same spectral type.    
Our measured rotation period and measured $v \sin i$ together imply that $R \sin i = 0.196 \pm 0.015 R_\odot$ for 2M0335+23, whose age of $24 \pm 3$ Myr is independently known. This is much larger than the radii of field M7 dwarfs, which have have $R = 0.12 R_\odot$ \citep{2014AJ....147...94D}. This result can be seen as either an independent confirmation of the evolutionary model prediction that young brown dwarfs are larger, or if the models are trusted, as independent
support for BPMG group membership of 2M0335+23.  Using the previously estimated radius of \citet{2015ApJS..219...33G}, we find that the inclination is $i = 54.4 \pm 6.6^\circ$.

\subsection{Flare Frequency Distribution}

\begin{figure*}
\plotone{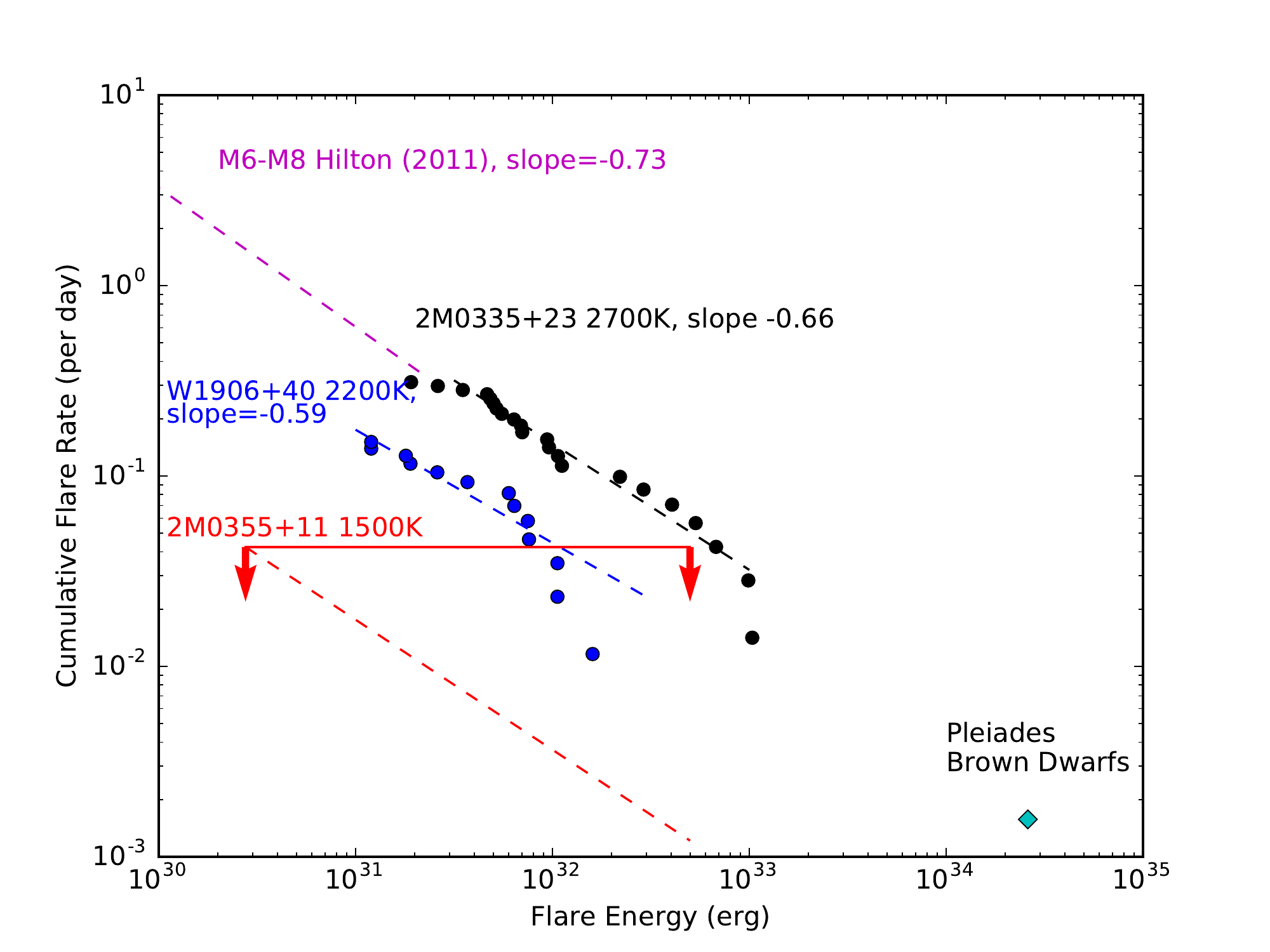}
\caption{The observed flare frequency distribution (FFD) and fit for the brown dwarf 2M0335+23 (black points and dashed line) compared to the L1 dwarf W1906+40 (blue) and upper limit for 2M0355+11 (red). The \citet{2011PhDT.......144H} fit to the  FFD for field M6-M8 dwarfs, transformed from $E_U$ to total energy as described in the text. \citet{2011PhDT.......144H} slope and normalization agree remarkably well with 2M0335+23's FFD measured at higher energies.\label{fig-ffd}}
\end{figure*}

Studies of flare stars have found that the cumulative flare frequency distribution (FFD) follows a power-law trend \citep{1972Ap&SS..19...75G,Lacy:1976qy}. We compute the cumulative frequency ($\nu$) of 2M0335+23 flares as the number ($N$) observed with a given energy or greater divided by the total time of observation (70.7 days) and plot the results in in Figure~\ref{fig-ffd}.  {\bf The statistical properties of power-law (``Pareto") distributions are reviewed by \citet{Arnold2015}.  We consider both the graphical technique of fitting a line to Figure~\ref{fig-ffd}
and maximum likelihood estimation: \citet{Arnold2015} notes that the two techniques are consistent and that the traditional graphical techinique is ``only slightly inferior" to the maximum likelihood estimates.  }
Considering flares in the energy range $4\times10^{31}$ erg to $1.1 \times 10^{33}$ erg, using frequency units of day$^{-1}$, and weighting each point by $\sqrt{N}$, we fit a linear relationship.

\begin{equation}
\log \nu = a +\beta \log(E/{10^{32} \rm{erg}})
\end{equation}

We find $\beta= -0.66 \pm 0.04$ and $a= -0.83 \pm 0.01$ for 2M0335+23.  An alternative expression of this power law dependence is:

\begin{equation}
dN \propto E^{-\alpha} dE
\end{equation}

Here $dN$ is the number of events between energy $E$ and $E+dE$.  As often discussed for flares, if $\alpha>2$ the total energy in small events (nanoflares) would diverge. Because $\beta = -\alpha +1$ 
(see \citealt{2014ApJ...797..121H} for helpful discussion), $\alpha=1.7$ for 2M0335+23.  We also plot limits
on the FFD for the L5$\gamma$ brown dwarf 2M0355+11. We also show the {\it Kepler} L1 dwarf W1906+40 \citep{2013ApJ...779..172G} FFD but over the energy range  $10^{31}$ erg to  $2 \times 10^{32}$ erg.  For this star, $a=-1.35\pm0.06$  and $\beta= -0.59 \pm 0.09$, but we caution that the slope depends sensitively on the energy range chosen.

{\bf An alternative approach is to use the maximum likelihood estimator for $\alpha$ \citep{Arnold2015}:

\begin{equation}
(\alpha -1) = n \left[ \sum_{i=1}^{n} \ln \frac{E_i}{E_{min}}\right]^{-1}
\end{equation}

where for 2M0335+23 case $n=19$ and $E_{min} = 4.66\times10^{31}$ erg. The uncertainty in this estimator can also be calculated (we used the software from \citet{2014PLoSO...985777A}). Because our sample is relatively small the uncertainty is large ($\pm 0.2$), and importantly, the estimator is biased. \citet{Arnold2015} shows it can be made unbiased by multiplying it by a factor of $\frac{n-2}{n}$, giving $\alpha = 1.79 \pm 0.21$. We conclude that the maximum likelihood estimate of $\alpha$ is consistent with the linear fit value. We adopt the (larger) uncertainty of $\pm 0.2$ from the maximum likelihood estimator. For W1906+40, the maximum likelihood estimator is $\alpha=1.6 \pm 0.2$.} 

The 2M0355+11 limit is shown in red in Figure~\ref{fig-ffd}. If one assumes that 2M0355+11 would follow the same power law slope as 2M0335+23 or W1906+40, then the red dashed-line upper limit shown in Figure~\ref{fig-ffd} applies. In any case, 2M0355+11's incidence of flares of $2 \times 10^{32}$ erg or greater is less than 13\% that of 2M0335+23 and less than 40\% that of the much older star W1906+40.  

\citet{2011PhDT.......144H} found that for four M6-M8 dwarfs, $a=20.49 \pm 3.3$ and $\beta = -0.73 \pm 0.1$ over the range $10^{27.94} \le E_U \le 10^{30.60}$ erg, where the frequency is in hour$^{-1}$, $E_U$ is the energy in U-band, and the reported intercept $a$ is at zero energy, not $10^{32}$ erg. In order to compare this FFD, we can make a crude energy correction by noting that \citet{Hawley:1991uq} found that the total energy of a \object{AD Leo} flare was $806/145=5.55$ times greater than the U-band energy (their Table 6).  (The 10,000K blackbody agrees well with this flare, see \citealt{2013ApJ...779..172G}).  Applying this correction, the \citet{2011PhDT.......144H} relation agrees remarkably well not only in slope but also normalization (Figure~\ref{fig-ffd}).  The agreement in normalization seems rather fortuitous given the uncertain energy corrections applied, the combination of multiple stars, and the fact that as a young brown dwarf, 2M0335+23 is both larger and more luminous than a field M7 dwarf.  

We conclude that the flares for ultracool dwarfs follow a power law over the range $5 \times 10^{28}$ to $10^{33}$ erg, a range of 4.3 orders of magnitude.  The existence of the Pleiades flare shows that this distribution must continue to at least $3 \times 10^{34}$, although we cannot determine whether it follows the same power-law slope or turns over.  The predicted rate using the extrapolated 2M0335+23 FFD for the Pleiades superflare is 1.4 per year. The striking aspect of this power-law slope is that it agrees so well not only with stellar M6-M8 dwarfs, but also ``inactive" M dwarf stars. It is also in excellent agreement with solar flares. 
\citet{1993SoPh..143..275C} found that $\alpha \approx 1.5$ for solar hard-X ray flares over three orders of magnitude.  \citet{2000ApJ...535.1047A} argue that solar flares follow a power low slope with $\alpha=1.8$ over the range $10^{24}$ to $10^{32}$ erg.  Power-law slopes near this value are explained by self-organized criticality models \citep{1991ApJ...380L..89L,2016SSRv..198...47A} where flares are due to ``avalanches" of many small magnetic reconnection events.   2M0355+11 may be at a temperature (1480K) below which fast magnetic reconnection events are no longer possible.

While the slopes ($\alpha$) are consistent between the brown dwarfs, the Sun, and other stars, the overall normalization of the flare rate ($a$) is much different.  
Flares on 2M0335+23 are much less frequent than in the rapidly rotating M4 dwarf GJ 1243 \citep{2016ApJ...829..129S}, comparable to ``inactive" M dwarfs studied by \citet{2014ApJ...797..121H}, and more  frequent than in the Sun.  It is perhaps most interesting to compare to the L1 dwarf W1906+40.  
At an energy of $10^{32}$ ergs, flares are 3-4 times as frequent on 2M0335+23 as on W1906+40. However, 2M0335+23 is 13 times more luminous than W1906+40, so despite its higher temperature, it is less efficient at converting its energy into flares; it also has seven times the surface area of W1906+40 so the flare rate per unit area is lower. It is intriguing that in the \citet{2010A&A...522A..13R} theory, W1906+40's predicted magnetic field is $3.1$ kG, 40\% stronger than 2M0335+23.

If we compute the total power in 2M0335+24 white light flares as the integral of $EdN$ from the solar microflare energy of  $10^{24}$ erg to the observed superflare energy of $2.6\times10^{34}$ erg using the 2M0335+23 FFD fit, we find  $\log L_{\rm WLF} / L_{\rm bol} = -4.2$ .  A more conservative upper limit $10^{33}$, and lower limit of $5 \times 10^{28}$ erg still gives $\log L_{\rm WLF} / L_{\rm bol} = -4.7$.

\subsection{The Pleiades Brown Dwarf Superflare}

With an energy of $2.6 \times 10^{34}$ erg, the CFHT-PL-17 flare event is comparable to superflares observed in {\it Kepler} G, K, and M stars \citet{2014ApJ...792...67C}, though below the mean observed superflare energy \citep{2016ApJ...829...23D}. It is helpful to consider the flare in terms of the observed peak absolute $M_{Kp} = 10.3$ (long cadence) or $\sim8.7$ (short cadence): If this brown dwarf were an unresolved companion to an A star ($M_{Kp} \approx 0$) or F dwarf ($M_{Kp} \approx 2.5$) the flare would be a detectable event with {\it Kepler}. However, the flare rates seen in some A and F stars by \citet{2012MNRAS.423.3420B} seem to be too high to be explained by brown dwarf companions, since they re-occur on timescales of 1-120 days whereas we expect less than one per year due to a brown dwarf.  X-ray triggered events have revealed that even more energetic superflares occur in young M dwarfs such as the $\sim30$-Myr old M dwarf binary \object{DG CVn} \citep{2016arXiv160904674O}. For comparison to other ultracool dwarfs, this superflare has more energy than  the L1 dwarf superflare we reported in Paper I of this series \citep{2017ApJ...838...22G} but less than the ASASAN-16ae L0 dwarf superflare \citep{2016ApJ...828L..22S}.  

As noted above, the extrapolated flare rate of 2M0335+23 suggests a superflare only $\sim1.4$ times per year.
We find that there are nine well-resolved Pleaides brown dwarfs in the spectral type range M6-M9 Campaign 4 for which we would have been detected a similar superflare. The combined superflare rate of these nine brown dwarfs is  $\sim1.7$ times per year. This suggests that the superflare may be understood as the high energy tail of the white light flare power-law distribution.  However, \citet{Rubenstein:2000nx} suggested superflares in solar-type stars may be the result of interactions with a planetary companion, and we have no information about whether CFHT-PL-17 has a lower-mass brown dwarf or exoplanet companion. {\bf We see no reason to invoke interactions with a substellar companion to explain the white light flares in either 2M0335+23 or CFHT-PL-17.}

\subsection{Complex flares: Is sympathetic flaring at work?}

The two 2M0335+23 complex flares (Fig~\ref{fig-flares}, Table whatever) are very well described as the sum of two individual flares that follow the template, {\bf and we assume through this section that they are occurring on 2M0335+23 rather than an unknown companion.}  The time separation is 14.6 minutes for the flare on mission day 2240 and 9.2 minutes for the flare on mission day 2287.  A third possible example of a complex flare occurs on mission day 2258, with a time separation of about 8 minutes. However, in the third case, the noise level is large enough that it is not altogether clear that two individual flares can be reliably identified. Thus, among the 22 flares illustrated in Fig.~\ref{fig-flares} , we can confidently state that about 10\% exhibit the occurrence of two flares within a time interval shorter than 20 minutes. 

Flares which occur within a short time interval of each other may belong to the class of ``sympathetic flares" (SF). By definition, SF are related to each other in the sense that a disturbance (of a kind that we will discuss below) generated by the first flare propagates to another active region and triggers a flare there.  However, a ``short" time interval between flares is not necessarily an indication of SF.  In fact, it may be difficult to identify with confidence a bona fide SF in certain stars. For example, a very active flare star may have multiple flares occurring randomly in multiple active regions within short time intervals, and these flares may have little or physical relationship to one another. Is there a way to distinguish between unrelated flares and SF? We suggest that one possible approach may be to consider the ratio of (i) the time interval T between two particular flares, and (ii) the mean time interval $T(m)$ between flares averaged over the length of the entire observing period.

For example, the M4 flare star GJ 1243 \citep{2014ApJ...797..122D} was observed by Kepler for a period of 11 months, during which 6107 ``unique events" were identified as flares. This star has an average of 18-19 flares per day, i.e. $T(m) = 75-80$ minutes. An example of a ``complex flare" is illustrated by \citet{2014ApJ...797..122D} in their Fig. 6, which spans an interval of 3.6 hours: 7 template flares are required to produce a good fit to the light curve. The average time interval between the template flares in this case is T = 31 minutes. This is already shorter than T(m) by a factor of 2, and might therefore suggest that SF could be at work. A fortiori, if we exclude one outlier flare at late times (with a peak at abscissa 549.865 days), inspection of their Fig. 6 suggests that 5 template flares occurred on GJ 1243 within a time interval of only 75 minutes, i.e. $T = 15$ minutes. This is 5-6 times shorter than $T(m)$, again suggestive of SF.

In contrast to the active flare star GJ 1243, when we return to considering the object of interest to us here (2M0335+23), we find that the flares in Fig.~\ref{fig-flares} were observed over a 66-day interval. This means that, with 21 flares in our sample, the mean interval between flares is $T(m) =  3$ days. The fact that we have identified two (or possibly 3) pairs of flares separated by only 20 minutes means that our pairs of flares have time separations T which are shorter than $0.01T(m)$. As a result, while we may assert that neighboring flares on 2M0335 are probably randomly related if they are separated by 3 days or more, it is much more difficult to make such an assertion for flares which are separated by less than 1\% of T(m). It seems more likely to consider the possibility the two flares which are separated by only 1\% of T(m) are physically related to each other. Specifically, is it possible that we are observing pairs of SF in 2M0335+23?

In the Sun, the SF possibility was subject to opposing claims in the 1930Õs based on optical data 
\citep{1937PASP...49..233R,1938ZA.....16..276W}. Conflicting claims for the existence (or non-existence) of SF surfaced again in the 1970Õs, based on radio and X-ray data (\citealt{1976SoPh...48..275F}: hereafter FCS). Based on X-ray data, FCS reported on the absence of significant evidence for SF except in one sub-set of their data: active regions which were closer to each other than a critical distance exhibited a 3.4$\sigma$ increase in the occurrence of ``short" time intervals between flares. Coincidentally, FCS defined ``short" as being $<20$ minutes, i.e. the same interval as we mentioned above in connection with flares on 2M0335+23. However, FCS seemed suspicious about even the one SF case they had detected because they could not identify ``any mode of propagation of a triggering agent in the solar atmosphere."  

In the case of stars, \citet{1971Ap......7...48O}[OT] analyzed the time intervals between successive flares in \object{YZ CMi} and \object{UV Cet} and found that the intervals in general followed a Poisson distribution.  However, in UV Cet, there exist certain ``sequences of closely spaced flares whose probability of occurrence is very small in the case of a Poisson process." In one case, 7 flares occurred in 5.4 minutes, and on each of 2 separate occasions, 3 flares were observed within a 2-minute interval. OT demonstrated that such small intervals of time between flares are highly improbable in the context of the overall Poisson distribution.  

Also in the case of the same flare star as that discussed by OT, \citet{1974A&A....37..219H}[HS] reported on an independent study of the time intervals between flares on UV Cet. In the course of 26 hours of observing, they detected 94 flares. Thus, they obtained T(m) = 17 min as the average time between flares. However, when they examined the distribution of time intervals between individual flares, the intervals ranged up to as long as 110 min. A Poisson distribution was found to fit the flare interval data at the 98\% confidence level with one proviso: only intervals larger than 4 min were included. At intervals shorter than 4 min, there is a large spike in the distribution: 38 of the 94 flares were found to have $T\le 4$ min. This excess at short times is far above what the Poisson distribution predicts. HS cited OT as having also ``noted" this excess at short times.  But then, with regards to the excess at short times, HS make the following statement (which has no explicit analog in OT): ``This might be due to triggering of the second flare by the first, like sympathetic flares on the sun."

A possible triggering agent for SF in the Sun was suggested by \citet{1968SoPh....4...30U}: a fast-mode MHD wave/shock which is launched into the corona by certain flares.  The idea is that as the wave/shock propagates through the corona, it may encounter a second active region: in that case, the wave/shock may perturb the second active region in such a way that a ``sympathetic" flare occurs in that active region. Evidence for disturbances propagating away from certain flare sites was at first based solely on chromospheric data, where a ``Moreton wave" was observed sweeping across the chromosphere. An observational break-through as regards a triggering agent for SF occurred with the launch of SOHO in 1995, when the Extreme-ultraviolet [EUV] Imaging Telescope (EIT) detected waves which propagate through large distances in the solar corona following certain events. The waves are referred to variously as ``EIT waves" or ``EUV waves." The waves were first interpreted as fast-mode MHD waves driven either by an erupting coronal mass ejection (CME) or as a blast wave driven by the energy release in a flare. The earliest data indicated that EUV waves propagate at speeds of 200-500 km/sec in the solar corona \citep{2000A&AS..141..357K}, but speeds as large as 1400 km/sec have been reported \citep{2013ApJ...776...58N}.  An extensive survey of multiple theories which have been proposed to explain EUV waves \citep{2016arXiv161105505L} has concluded that the waves in the Sun are ``best described as fast-mode large-amplitude waves or shocks that are initially driven by the impulsive expansion of an erupting CME in the low corona."

In the case of stellar (and brown dwarf) flares, is it possible that we might rely on solar-like phenomena to understand SF? Flares on stars involve magnetic energy release, so to the extent that a stellar flare contributes to the launch of a blast wave in the corona (analogous to the Sun), we may expect that flare-induced EUV waves could contribute to stellar SF. What about EUV waves generated by CMEÕs? Can we count on those to occur in flare stars and serve to launch EUV waves to help generate SF? Although flares and CMEÕs in the Sun both involve release of magnetic energy, they do not always occur together: one can occur without the other, depending on local details of the parent active region. We note that at least one detection of a stellar CME has been reported from an active K dwarf which is known to be a flare star \citep{2001ApJ...560..919B}.  

Let us examine the hypothesis that the 2 complex flares which we have detected in 2M0335+23 involve SF which are triggered by the equivalent of an EUV wave. In this context, the maximum speed of the wave would be obtained if the two active regions which are involved in the individual flares were located at antipodal points on the surface of 2M0335+23, at a distance $\pi R$ from each other. Inserting the radius $R$ of 2M0335+23, and inserting time delays of 14.6 and 9.2 minutes, the SF hypothesis leads to $v(EUV) < 600$ and $950$ km s$^{-1}$. Such values are well within the range of EUV wave speeds which have been reported in the solar corona. With fast-mode speeds determined mainly by the Alfven speed (which greatly exceeds the thermal speed of order 100-200 km/sec in a 1-2 MK corona), our SF interpretation suggests that Alfven speeds in the corona of 2M0335+23 may not differ greatly from those in the solar corona. In the latter, a map of coronal Alfven speeds reported by \citet{2015ApJ...812..119S} spans a range from 500 to 900 km/sec at essentially all latitudes within a radial distance of 5 solar radii.

\subsection{Complex flares: Does the weakness of the second flare contain physical information?}
 
We note that for the complex flares in Figure~\ref{fig-flares}, when the tail of the first light curve is subtracted from the second flare, the amplitude at the peak of the second flare is smaller than the amplitude at the peak of the first flare. We ask: Is this ``weaker secondary" a common feature of complex flares?
 
To address this, we consider some flare data which were recorded in different settings.
 
\begin{enumerate}
  
\item	A large homogeneous sample of X-ray flares which were recorded by Chandra in the Orion Ultradeep Project \citep{2008ApJ...688..418G}. In their study of the 216 brightest flares from 161 pre-main sequence stars, 8 events were classified as ÒdoubleÓ flares, i.e. they Òlook like two overlapped typical flaresÓ. By subtracting off the tail of the first flare in each case, we evaluated the ratio of peak 2 to peak 1, and we found the following values: 0.3, 0.4, 0.2, 0.5, 1.1, 0.2, 0.3, and 0.1. Thus, in 7 out of 8 cases, the Òsympathetic flareÓ has a smaller amplitude than the original flare.
  
\item	Optical data (in the $r$ band) for stars in the intermediate age cluster M37 (0.55 Gyr) resulted in detection of several hundred flares from cluster members \citep{2015AJ....150...27C,2015ApJ...814...35C}. \citet{2015ApJ...814...35C} drew attention to the result that their Òalgorithm often detects secondary flares which occur during the decay of a much larger flareÓ. Visual inspection of Fig. 2 in \citet{2015ApJ...814...35C} and Figs. 17, 18 in \citet{2015AJ....150...27C} suggests that as many as 8 or 9 secondary flares can be identified among the plotted light curves of 23 stars: in all cases but one, the flare which occurred later in time was smaller in amplitude than the first flare. 
  
\item A small sample (15-20) of light curves in a variety of visible and near UV wavelengths from a number of solar neighborhood flare stars has been presented by \citet{Gershberg:2005lr}[,pp. 195-205] 
Secondary flares can be identified in at least 5 cases, and in all cases, the flare which occurred later in time had a smaller peak intensity. These examples suggest that, no matter which wavelength range we examine, the flare which occurs later in time (the ``secondary") in a ``close double" has (in most cases) a smaller amplitude than the flare which occurs earlier. We believe that this is a feature which contains information related to one of the key physical processes involved in sympathetic flaring.
 
\end{enumerate}
 
 How might this ``weaker secondary" behavior be understood in the context of the SF explanation proposed above (i.e. SF is triggered by a fast-mode MHD wave)? We suggest that it may be understood in terms of wave refraction. When fast-mode MHD waves propagate through an inhomogeneous medium, the compressive nature of the waves has the effect that the waves are refracted away from regions of high Alfven speed ($v_A$), and are refracted into regions where $v_A$ is small \citep{1968SoPh....4...30U}. (Alfven waves, lacking compression, refract differently.) Tests of UchidaÕs predictions have been provided by observations (e.g. Shen et al. 2013) and modelling (e.g,  \citealt{1980ApJ...241.1186S},\citealt{2016ApJ...820...16J}) of fast-mode waves propagating through a variety of structures in the solar corona. However, the tests mentioned in the preceding sentence were based on indirect inferences of the $v_A$ value in different regions of the Sun. For more reliable tests of UchidaÕs theory, it is preferable to consider a medium where in situ measurements of field strength and ion density can be made directly: in such a medium, the local value of $v_A$ can be calculated,  various wave modes can be distinguished (fast MHD, slow MHD, Alfvenic), and UchidaÕs predictions can be tested directly. The solar wind is one such medium. In the solar wind at a radial distance of about 1 AU, data from the ACE satellite have been used to demonstrate that fast mode waves are indeed depleted in {\bf high-$v_A$ regions \citep{2001JGR...10618625S}}, and fast mode waves are indeed enhanced in regions of low-$v_A$ \citep{2003ApJ...583..496M}.
 
Now let us consider how these results might find an application in SF in low-mass stars. Suppose an initial flare is triggered (somehow) in a certain active region (AR), thereby launching a fast-mode EUV wave with a certain speed: the speed will be related to the $v_A$ value in the AR where the flare was initiated.  Suppose there are two other ARÕs on the surface of the star, AR-A with large $v_A$, and AR-B with small $v_A$. What happens when the EUV wave approaches AR-A? The fast-mode wave will be refracted away from AR-A because of the locally large $v_A$ : the wave will be unable to penetrate into AR-A. Therefore, a SF is unlikely to occur in AR-A. But in the case of AR-B, the fast-mode wave will be refracted into the AR, thereby having a chance to perturb the plasma inside AR-B, perhaps enough to initiate a flare. In this scenario, a SF is more likely to occur in an AR with a small value $v_A$: or, in the terminology of \citet{2016ApJ...820...16J}, the ``impact" of the wave on AR-B (with its lower $v_A$ ) would be larger. What might cause the $v_A$ value in AR-B to be smaller? It could be either weaker field or higher density, or both. In cases where the field is weaker, we expect that (other things being equal) a flare in such an AR will have (in general) a smaller total energy. Thus the flare originating in AR-B (where the ``impact" of the fast-mode wave is largest) will be ``weaker," and is expected to have a smaller peak amplitude. If this is a correct interpretation of SF in stars, then the phenomenon of the ``weaker secondary" may provide an observational signature of the physics of refraction of fast-mode MHD waves in the inhomogeneous corona of a flare star.

\section{Conclusions}

White light flares on a young (24 Myr) M7-type brown dwarf are similar in most respects to stellar M dwarf flares, including their light curves, power-law flare frequency distribution, and sympathetic flaring.  Adding a flare on a Pleiades brown dwarf, we see that these flares extend up to at least $2.6 \times 10^{34}$ erg. However, we observe no white light flares on the L$5\gamma$ brown dwarf despite its known young age and rapid rotation.
Since there is overwhelming observational and theoretical evidence that magnetic fields exist on L and T-type brown dwarfs, we conclude that the change in flare rates is direct evidence that fast magnetic reconnection is suppressed or forbidden at temperatures $\sim1500$K.  In this work, we have studied brown dwarfs of known age. In our next paper, we will measure the white light flare rates of a sample of field late-M and L dwarfs and investigate their dependence on age, rotation, effective temperature, and other observable properties.

\acknowledgments

We thank James Davenport and Rachel Osten for discussions of stellar flares, Jonathan {Gagn{\'e}} and Jackie Faherty for  discussions of moving groups,  Mike Liu and Conard Dahn for comments on the preprint, and the anonymous referee and statistical consultant for suggestions.
This paper includes data collected by the Kepler mission. Funding for the Kepler mission is provided by the NASA Science Mission directorate. The material is based upon work supported by NASA under award Nos. NNX15AV64G, NNX16AE55G, and NNX16AJ22G.  A.J.B. acknowledges funding support from the National Science Foundation under award No.\ AST-1517177. 
Some of the data presented in this paper were obtained from the Mikulski Archive for Space Telescopes (MAST). STScI is operated by the Association of Universities for Research in Astronomy, Inc., under NASA contract NAS5-26555. Support for MAST for non-HST data is provided by the NASA Office of Space Science via grant NNX09AF08G and by other grants and contracts.
Some of the data presented herein were obtained at the W.M. Keck Observatory, which is operated as a scientific partnership among the California Institute of Technology, the University of California and the National Aeronautics and Space Administration. The Observatory was made possible by the generous financial support of the W.M. Keck Foundation. 
This research has made use of NASA's Astrophysics Data System, the VizieR catalogue access tool, CDS, Strasbourg, France, and the NASA/ IPAC Infrared Science Archive, which is operated by the Jet Propulsion Laboratory, California Institute of Technology, under contract with NASA.  We have also made use of the 
\dataset[List of M6-M9 Dwarfs]{https://jgagneastro.wordpress.com/list-of-m6-m9-dwarfs/} maintained by Jonathan {Gagn{\'e}}.

\software{IRAF, AstroPy \citep{2013A&A...558A..33A}, photutils, emcee \citep{2013PASP..125..306F}, PyKE \citep{2012ascl.soft08004S}, APLpy, powerlaw \citep{2014PLoSO...985777A} }


\facility{Kepler, Keck:II (NISPEC)}

\bibliography{../astrobib}


\end{document}